\documentclass[prx,twocolumn,superscriptaddress,nofootinbib,noshowpacs,preprintnumbers,longbibliography,floatfix]{revtex4-2}
\usepackage[utf8]{inputenc}
\usepackage{graphicx}
\usepackage{float}
\usepackage{amssymb}
\usepackage{amsmath}  
\usepackage{adjustbox}
\usepackage{mathtools}
\usepackage{dsfont}
\usepackage{array}
\usepackage{bm,fixmath}
\usepackage{mathrsfs}
\usepackage{pifont}
\usepackage{multirow}
\usepackage{upgreek}
\usepackage{xcolor}
\usepackage{bm}
\usepackage{bbm}
\usepackage{physics}
\usepackage{comment}
\usepackage{tikz}
\usepackage[normalem]{ulem}
\usepackage{qcircuit}
\usepackage{color, soul}
\usepackage{placeins}
\usepackage{multirow}
\usepackage{booktabs}
\usepackage[pdftex,
            pdftitle={QCNN fermion scattering},
            pdfauthor={},
            bookmarks,
            colorlinks,
            linkcolor=myblue,
            citecolor=mymagenta,
            menucolor=black,
            urlcolor=myblue,
            plainpages=false,
            pdfpagelabels,
            hypertexnames=false]{hyperref}
\usepackage{orcidlink}
\usepackage[percent]{overpic}

\graphicspath{{figures/}}

\definecolor{mymagenta}{RGB}{200, 0, 100}
\definecolor{myblue}{RGB}{45, 48, 146}
\definecolor{mypurple}{RGB}{200, 112, 255}

\begin{document}

\title{Quantum Machine Learning for particle scattering entanglement classification}

\author{Hala Elhag\, \orcidlink{0000-0003-1638-8022}}
\email{hala.elhag@desy.de}
\affiliation{Deutsches Elektronen-Synchrotron DESY, 15738 Zeuthen, Germany} 

\author{Yahui Chai\,}
\email{yahui.chai@desy.de}
\affiliation{Deutsches Elektronen-Synchrotron DESY, 15738 Zeuthen, Germany}

\date{\today}

\begin{abstract}

Entanglement is a key quantity for characterizing quantum correlations in particle scattering processes, but its direct evaluation is computationally demanding on quantum hardware. In this work, we investigate whether fermion density profiles, which are easier to access, can serve as proxies for entanglement by framing the problem as a classification task across multiple entanglement thresholds. Using the fermion scattering in the Thirring model as a test bed, we compare Quantum Convolutional Neural Networks (QCNNs) with classical CNNs of comparable parameter counts, and find that QCNNs achieve consistently competitive or superior accuracy with faster convergence and lower variance. Notably, we observe that increasing the model size does not improve the performance within the architectures studied here, and larger models appear to be more sensitive to the choice of encoding. Instead, a compact 4-qubits QCNN provides the best results, suggesting the importance of trainability and encoding choices over model scaling. These findings demonstrate the potential of quantum and quantum-inspired machine learning models for extracting nontrivial quantum information from accessible observables, with implications for high-energy physics and quantum many-body systems.

\end{abstract}

\maketitle

\section{Introduction}

As an application for the use of quantum computing in high-energy physics (HEP), Quantum Machine Learning (QML) has gained increasing interest as a potential tool for tackling complex physical problems~\cite{Guan:2020bdl}. The ability of quantum systems to naturally encode and process high-dimensional, entangled information makes QML particularly appealing for applications in HEP, where many phenomena are inherently quantum in nature and computationally demanding to analyze using classical methods. As a result, QML has begun to establish itself as a promising framework for modeling, classification, and feature extraction in particle physics and related fields~\cite{farhi2018classification}.

Despite its potential, QML is still in its early stages of development and lacks a universally accepted or standardized structure. Unlike classical machine learning, where well-established architectures such as convolutional neural networks (CNNs) and transformers dominate, QML remains a rapidly evolving domain with diverse and often problem-specific designs. This nascent state of the field presents both challenges and opportunities. While it complicates direct benchmarking and comparison, it also allows for innovative exploration of novel quantum-enhanced architectures tailored to specific scientific problems. 

Encouragingly, early theoretical and experimental results in QML have demonstrated promising performance across a range of tasks~\cite{Mitarai_2018_paramShift,farhi2018classification,Benedetti:2019inj,PhysRevA.101.032308,elhag,Guan:2020bdl,felser2021quantuminspiredmachinelearninghighenergy,Araz_2021,Araz_2022,Gianelle_2022,Belis_2024}, suggesting that quantum and quantum-inspired models could lead to improved architectures and enhanced predictive capabilities. These initial successes motivate further investigation into how quantum properties such as superposition and entanglement can be harnessed to extract meaningful physical insights, particularly in domains where classical methods face fundamental limitations.

A natural motivation for the present work comes from quantum simulations of particle scattering processes~\cite{chai2025, chai2025fermionhardware, Chai2025fermionicwavepacket, Davoudi2024, davoudi2025, Roland_2024, farrell2025, Zemlevskiy2025, zemlevskiy2026, Julian_2025}. In such dynamics, local observables such as the fermion density describe how particles redistribute in space and time, while entanglement entropy characterizes the nonlocal quantum correlations generated during the collision. The latter contains physically valuable information, but its direct evaluation is considerably more demanding, especially for large many-body systems and on quantum devices. By contrast, the fermion density is much easier to access. This raises a natural question: can fermion density profiles serve as useful proxies for identifying scattering events that generate substantial entanglement?

In this work, we address this question through an entanglement-threshold classification task. Rather than predicting the exact entanglement value, we ask whether the entanglement associated with a given scattering event exceeds a chosen threshold. This provides a coarse-grained but physically meaningful way to distinguish different correlation regimes, and serves as a practical first step toward inferring expensive quantum diagnostics from accessible observables.

We further study this problem using QML models in direct comparison with classical machine learning baselines, in order to assess whether quantum structured architectures can offer improved accuracy while remaining efficient. As a concrete testbed, we consider fermion--antifermion scattering in the one-dimensional massive Thirring model and use the resulting fermion density profiles as input data. Beyond model comparison, this setting also provides a useful step toward future applications involving genuinely quantum data in high-energy physics and quantum many-body systems.

Among QML architectures, the Quantum Convolutional Neural Network (QCNN)~\cite{cong} is a particularly appealing candidate for the present task. QCNNs have shown promising performance in several classification problems~\cite{elhag, Tak, Samuel}, and their hierarchical structure can improve trainability compared with more generic variational circuits. In addition, although small QCNNs are efficiently classically simulable, they can still be viewed as useful quantum-inspired architectures that retain the structural features of quantum models. For these reasons, we employ QCNNs to perform the fermion scattering entanglement-threshold classification task and compare their performance with classical CNN baselines.

\section{Fermion Scattering Background}
In this work, we study fermion–antifermion scattering in the one-dimensional massive Thirring model~\cite{Thirring1958}, which
provides a simple interacting setting for investigating real-time quantum dynamics. Using the Kogut–Susskind
staggered formulation~\cite{Kogut1975, Susskind1977}, the lattice Hamiltonian is given by~\cite{Mari_Carmen}
\begin{equation}
    \begin{aligned}
    H &= \sum_{n}\left(\frac{i}{2a} \left( \xi_{n+1}^{\dagger}\xi_{n} - \xi_n^{\dagger}\xi_{n+1}\right) + (-1)^n m \ \xi_n^{\dagger}\xi_n \right)\\
    &+ \sum_{n} \frac{g}{a} \xi_n^{\dagger}\xi_n \xi_{n+1}^{\dagger} \xi_{n+1} \, ,    
    \end{aligned}
    \label{eq:hamiltonian_lattice_fermionic}
\end{equation}
where $\xi^{\dagger}_n$ and $\xi_n$ are fermion creation and annihilation operators; $a$ is the lattice spacing, $m$ is the fermion mass, and $g$ denotes the four-fermion interaction term. Without loss of generality, we set $a=1$ for the rest of this work. 

Following Ref.~\cite{Chai2025fermionicwavepacket, chai2025fermionhardware}, we investigate the scattering between a fermion and an anti-fermion wave packet for the interacting case. As in the reference, we prepare the initial scattering state by creating a fermion and an antifermion wave packet
on top of the vacuum $\ket{\Omega}$, 
\begin{equation}\label{eq:initial_state}
    \ket{\psi(t=0)} = D^{\dagger} C^{\dagger} \ket{\Omega},
\end{equation}
where $C^{\dagger} $  and $D^{\dagger}$ are creation operators for fermion and anti-fermion wave packets, respectively. These operators can be expressed as the linear combinations
 \begin{equation}
    \begin{aligned}
        C^{\dagger}(\phi^{c}) &= \sum_k \phi_k^c c_k^{\dagger} = \sum_n \tilde{\phi}_n^c \xi^{\dagger}_n, \\
        D^{\dagger}(\phi^d) &= \sum_k \phi_k^d d_k^{\dagger} = \sum_n \tilde{\phi}_n^d \xi_n,
    \end{aligned}
    \label{eq:creation_operators_gaussian_particle}
\end{equation}
the operator $c_k^{\dagger}, d_k^{\dagger}$ are fermion and antifermion creation operators with specific momentum $k$, as defined in Eq.~A6 in Ref.~\cite{Rigobello2021}, and the Gaussian coefficients $\phi_k^{c(d)}$ in momentum space are given by
\begin{equation}
    \phi_k^{c(d)} =  \frac{1}{\sqrt{\mathcal{N}_k^{c(d)}}} e^{-ik\mu_n^{c(d)}} e^{-(k-\mu_k^{c(d)})^2 / 4\sigma_k^2}\, .
    \label{eq:coeff_momentum}
\end{equation}
In above $\mu_n^{c(d)}$ corresponds to the central position of the wave packet, $\mu_k^{c(d)}$ to the mean momentum, $\sigma_k$ represents the width in momentum space, and $\sqrt{\mathcal{N}_k^{c(d)}}$ is a normalization factor. The coefficients in position spapce $\tilde{\phi}_n^{c(d)}$ can be obtained by fourier transformation of the $\phi_k^{c(d)}$

The time-evolved scattering state is then given by
\begin{equation}
    \ket{\psi(t)} = e^{-iHt}\ket{\psi(0)},
    \label{eq:time_evolved_state}
\end{equation}
where \(H\) is the interacting Hamiltonian in Eq.~\eqref{eq:hamiltonian_lattice_fermionic}. 
From this state, one can compute observables such as the excess local fermion density above the vacuum,
\begin{equation}
    \Delta \langle \xi_n^\dagger \xi_n \rangle  = \bra{\psi(t)} \xi_n^\dagger \xi_n \ket{\psi(t)} - \bra{\Omega} \xi_n^\dagger \xi_n \ket{\Omega},
    \label{eq:fermion_density}
\end{equation}
which provides a direct characterization of the scattering dynamics in real space and time.

To quantify entanglement generation during the scattering process, we consider the bipartite entanglement entropy. 
For a bipartition of the lattice into subsystems \(\mathcal{L}_n = \{l < n\}\) and \(\mathcal{R}_n = \{l \geq n\}\), the reduced density matrix of subsystem \(\mathcal{L}_n\) is
\begin{equation}
    \rho_n(t) = \Tr_{\mathcal{R}_n} \left[\ket{\psi(t)}\bra{\psi(t)}\right],
    \label{eq:reduced_density_matrix}
\end{equation}
and the corresponding von Neumann entanglement entropy is
\begin{equation}
    S_n(t) = - \Tr \left[\rho_n(t)\log \rho_n(t)\right].
    \label{eq:entropy}
\end{equation}
In this work, we are interested in the entanglement generated across different bipartitions during the fermion--antifermion collision. More specifically, we consider the excess entanglement entropy relative to the vacuum,
\begin{equation}
    \Delta S_n(t)=S_n(t)-S_n^{\mathrm{vac}},
\end{equation}
where $S_n^{\mathrm{vac}}$ denotes the vacuum entanglement entropy for the same bipartition.

While the fermion density in Eq.~\eqref{eq:fermion_density} is relatively accessible through the measurement of a local operator, the direct evaluation of entanglement entropy is considerably more demanding, especially on quantum hardware. Indeed, estimating \(S_n(t)\) generally requires access to the reduced density matrix or additional tomography-like procedures, whose cost increases rapidly with system size. This motivates our strategy: instead of directly computing the entanglement entropy, we use fermion density profiles in the scattering process as input data and train machine learning models to classify whether the excess entanglement entropy exceeds a given threshold.

The dataset used in this work is generated from tensor-network simulations of fermion--antifermion scattering in the 40 site massive Thirring model. We consider
\begin{equation}
    m,g \in \{0.1,0.2,\ldots,0.9\},
\end{equation}

and momentums of wavepackets with
\begin{equation}
\mu_k^{c} \in \{1,2,3,4,5,6\}, \qquad \mu_k^{d} \in \{-1,-2,-3,-4,-5,-6 \}.
\end{equation}
For each parameter choice, we compute the real-time evolution and extract both the fermion density profiles and the corresponding excess bipartite entanglement entropy.

For the supervised classification task, the binary label is defined from a reference entanglement value evaluated at a time identified from the density evolution. For each scattering event, we scan the time-dependent fermion density profile and determine the first time \(t^\ast\) at which the fermion and antifermion wave packets are well separated after the collision, which we define operationally as the point where the density maximum and minimum are more than 20 lattice sites apart. We then define the entanglement indicator as the excess central bipartite entropy
\begin{equation}
    \Delta S_{\mathrm{mid}}(t^\ast)=\frac{\Delta S_{19}(t^\ast)+\Delta S_{20}(t^\ast)}{2},
\end{equation}
and assign the label according to whether \(\Delta S_{\mathrm{mid}}(t^\ast)\) is above or below a chosen threshold \(S_{\rm th}\).

\begin{figure}
    \centering
    \includegraphics[width=\linewidth]{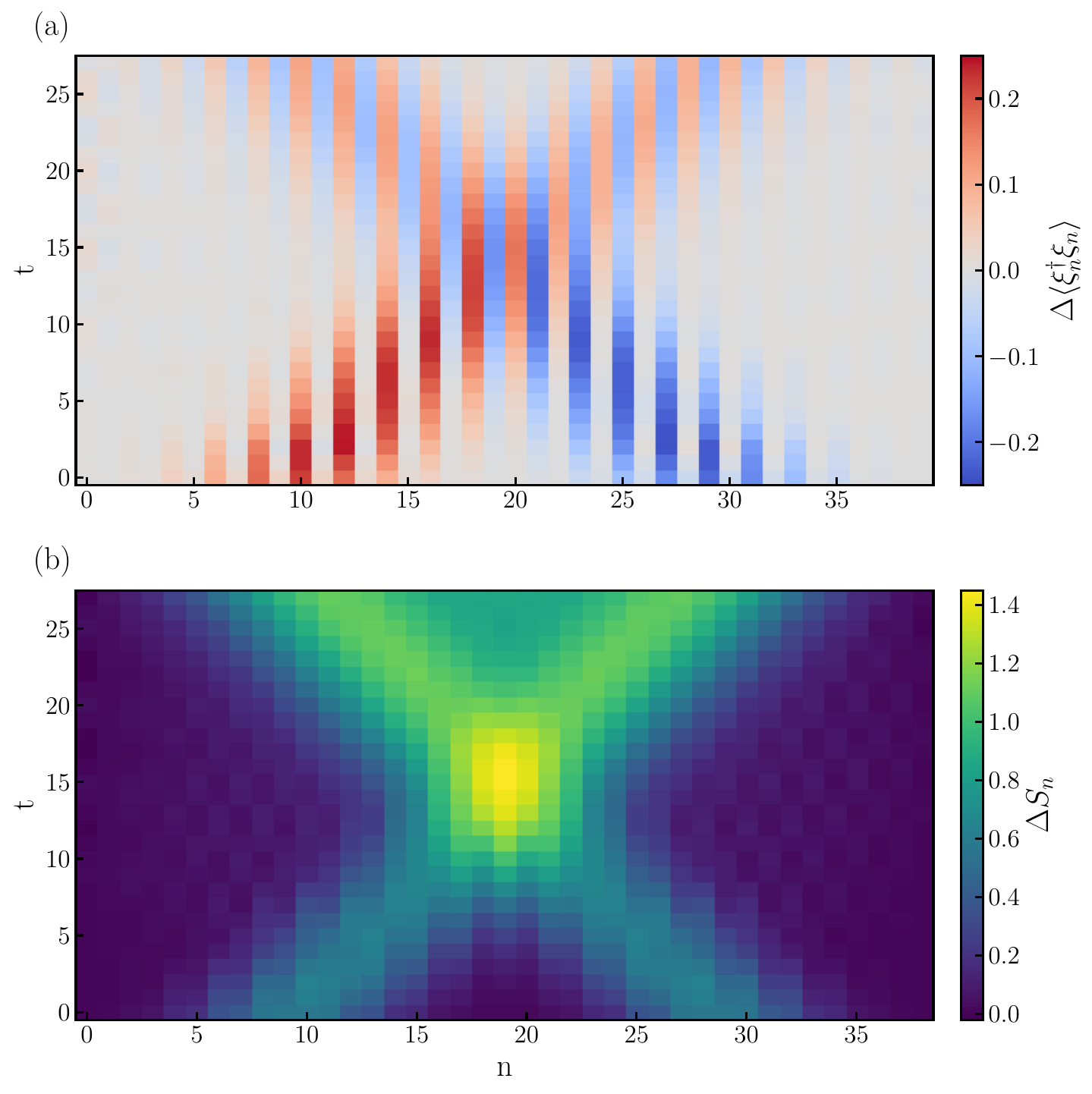}
    \caption{The fermion-antifermion scattering process in a 40 sites system with $(m,g)=(0.4, 0.5)$. (a) The excess local fermion density above the vacuum. (b) The excess bipartite entanglement entropy relative to the vacuum. The $x$-axis is the site index, and the $y$-axis represents time. }
    \label{fig: entropy}
\end{figure}

\section{Model and methods}

\begin{figure}
    \centering
    \includegraphics[width=\linewidth]{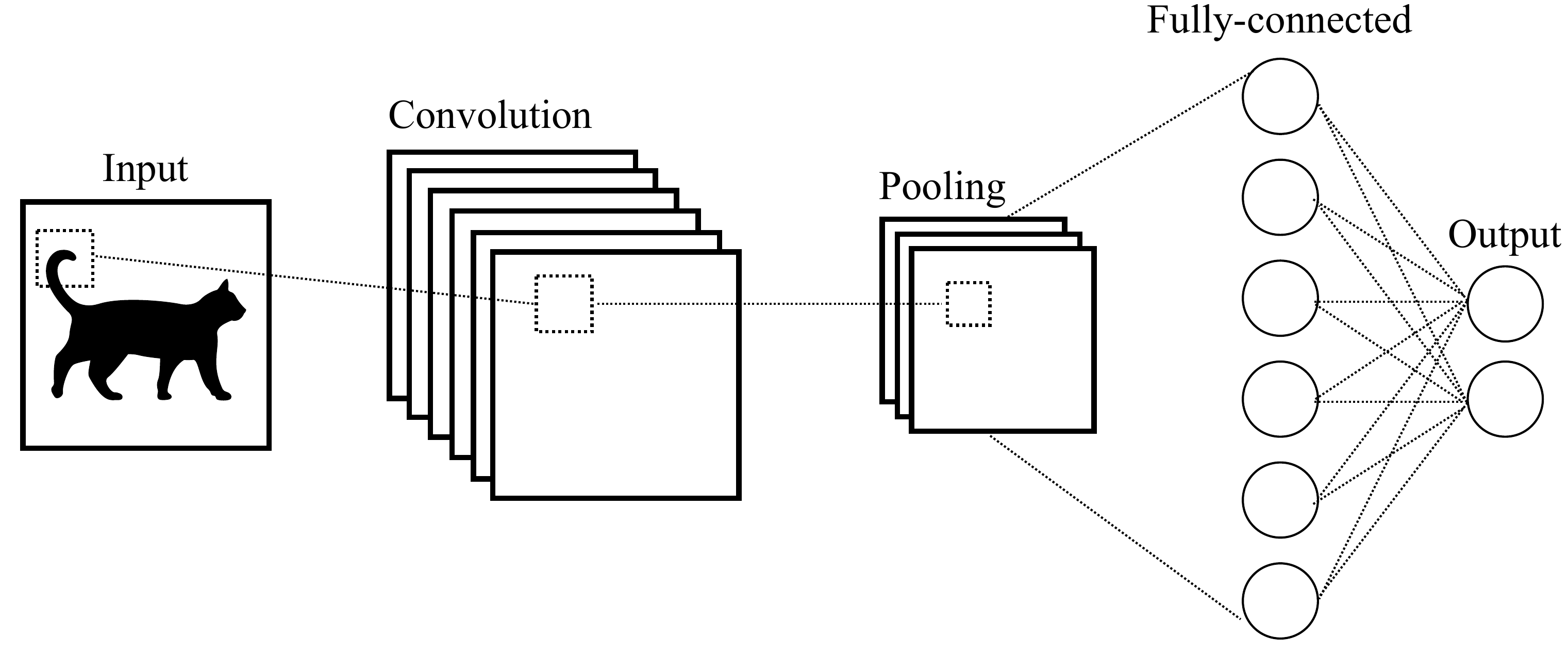} 
    \caption{A general CNN structure~\cite{elhag}.}
    \label{fig:cnn}
\end{figure}

\begin{figure}
    \centering
    \includegraphics[width=\linewidth]{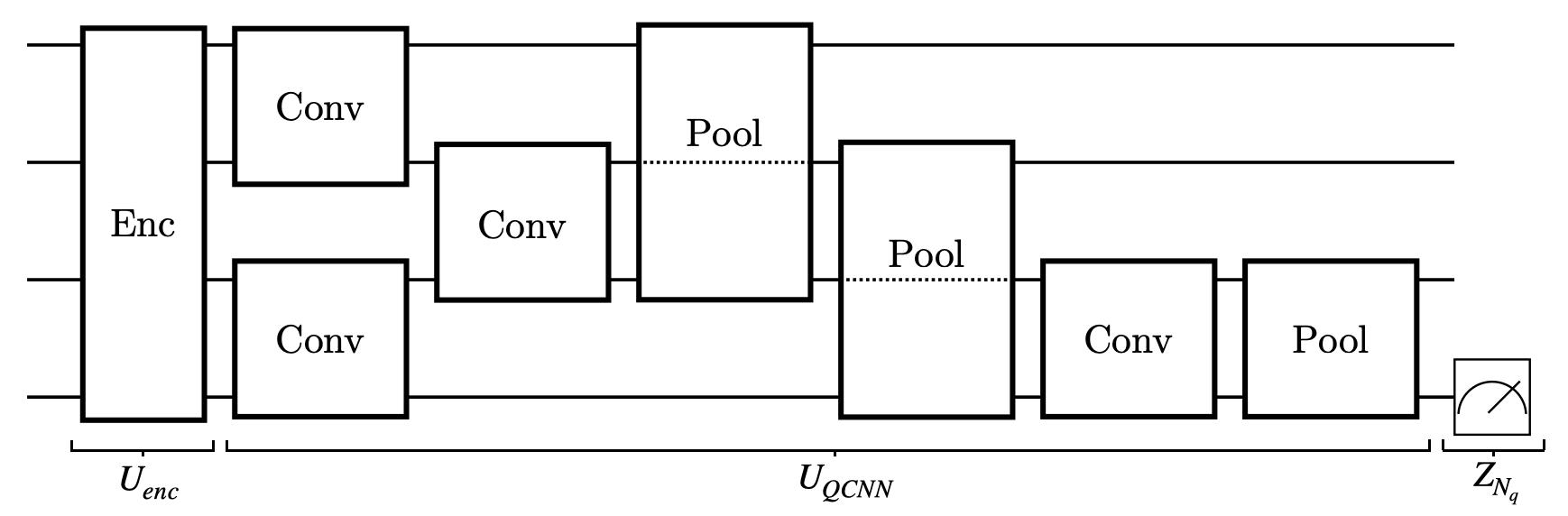} 
    \caption{Basic structure of a QCNN model~\cite{elhag}.}
    \label{fig:qcnn}
\end{figure}

The general structure of a CNN model can be found in Fig.~\ref{fig:cnn}. The network is primarily characterized by convolutional blocks, which convolve over an input image to extract features, and a pooling block, which reduces the dimensionality by emphasizing the most important features among a group of input values output from a convolutional block. The QCNN model is inspired by this structure and mainly follows the architecture presented in~\cite{elhag}. This architecture is described in FIG. \ref{fig:qcnn}, where the convolutional block is built based on the $SU(4)$ two-qubit unitary circuit containing 15 trainable parameters and three CNOT gates. The pooling block has 9 trainable parameters and one CNOT which reduces the circuit dimensionality by disregarding the output from some qubits and passing the output from other qubits such as the way displayed in the figure. 

Mainly, three QCNN models are to be investigated here. These are 4, 8, and 16-qubits system models. The 4-qubits model incorporates 48 trainable parameters in total, the 8-qubits model has 72 parameters and the 16-qubits model has 96 parameters. For each model, the dataset is preprocessed accordingly to match the number of qubits as the input data. In that manner, PCA is used to reduce the dimensionality to 4, 8, and 16 components for each model. The types of encoding used are Hardware Efficient Embedding (HEE), and Tensor Product Embedding (TPE) for comparison purposes. These encoding types were studied more intensively in~\cite{elhag} and~\cite{2021arXiv211014753T}. The QCNN models are compared to two CNN models, one with 51 trainable parameters and a larger model composed of 113 trainable parameters. 

\section{Results}

The classification task was performed on fermion scattering density images using different entanglement threshold values. The main goal is to determine the amount of entanglement found in a fermion scattering after the collision of fermion and antifermion. To this end, each classification is made for two classes corresponding to entanglement values above and below a given threshold. Performing this task on several threshold values would indicate an entanglement range for which a given fermion scattering event would lie on. Ultimately, one could repeat the task for much shorter ranges to obtain an accurate entanglement amount for the particle scattering event of interest, which could also be extended to a multiclass classification task.   

\begin{figure}[h!]
    \centering
    \includegraphics[width=\linewidth]{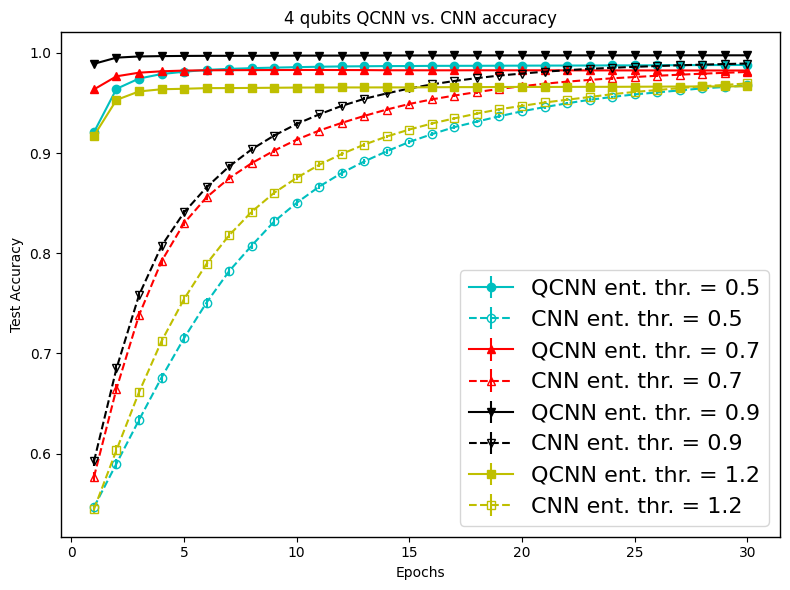}
    \caption{Test accuracy results averaged over $10^3$ runs of the 4-qubits HEE QCNN compared to CNN for the different classification thresholds.}
    \label{fig:qcnn_cnn_4q}
\end{figure}

The classification accuracy results obtained for four different entanglement thresholds are summarized in Table~\ref{tab:qcnn_cnn_results} and shown in Fig.~\ref{fig:qcnn_cnn_4q}. The QCNN model employs HEE encoding and was trained with 48 parameters, while the CNN model has 51 trainable parameters. For both models, the Adam optimizer~\cite{2014arXiv1412.6980K} was used to update the parameters with the mean squared error (MSE) loss function. 

For all threshold values, the QCNN model converges faster and generally outperforms its corresponding CNN model, even at later epochs. An exception occurs for thresholds 0.7 and 1.2, where both models achieve comparable performance during the final training stages. From Table~\ref{tab:qcnn_cnn_results}, it can be observed that the QCNN consistently achieves higher or comparable accuracy with lower variance across most thresholds, particularly for intermediate thresholds such as 0.5 and 0.9. At threshold 0.9, for example, the QCNN reaches a peak accuracy of 99.76\%, exceeding the CNN performance by a noticeable margin.

The number of images, (\textbf{n. images}) shown in Table~\ref{tab:qcnn_cnn_results}, per threshold category varies, to match a balanced number for each of the classes within each threshold, and 20\% of them are used for test. This reflects differences in the underlying data distribution. Despite this, the QCNN maintains strong performance, indicating robustness to dataset size variations. However, at threshold 0.9, where the number of images is the highest, both models achieve their best performance compared to other thresholds with fewer samples. This suggests that a larger dataset improves classification accuracy and highlights the importance of data availability in enhancing model performance, particularly for capturing more complex features in the distribution.

Overall, these results indicate the effectiveness of the QCNN architecture in capturing relevant features of the fermion scattering density data with fewer parameters and faster convergence. However, the dataset used to obtain the results are heavily suppressed by PCA reduction to four components only. This motivates the exploration of more advanced models with higher scaling, as well as alternative encoding strategies. 

\begin{table}[h]
\centering
\begin{tabular}{|c|c|c|c|}
\hline
\textbf{Ent. Thr.} & \textbf{n. images} & \textbf{QCNN Acc (\%)} & \textbf{CNN Acc (\%)} \\
\hline
0.5 & 806 & 98.80 $\pm$ 0.02 & 96.74 $\pm$ 0.12 \\
\hline
0.7 & 1516 & 98.24 $\pm$ 0.02 & 98.10 $\pm$ 0.13 \\
\hline
0.9 & 2314 & 99.76 $\pm$ 0.01 & 98.94 $\pm$ 0.12 \\
\hline
1.2 & 1116 & 96.67 $\pm$ 0.03 & 96.96 $\pm$ 0.11 \\
\hline
\end{tabular}
\caption{Test accuracy averaged over $10^3$ runs at epoch 30 for the 4-qubits QCNN model with 48 parameters compared to the CNN model with 51 parameters.}
\label{tab:qcnn_cnn_results}
\end{table}

\begin{figure}[h!]
    \centering
    \includegraphics[width=\linewidth]{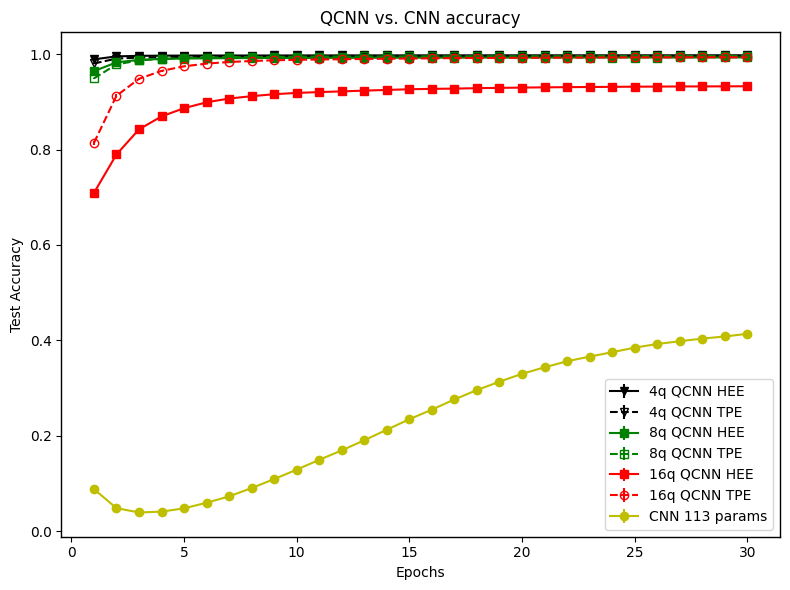}
    \caption{Test accuracy results averaged over $10^3$ runs of the 4-qubits, 8- and 16-qubits QCNN model with HEE and TPE for the classification threshold of 0.9 compared to the CNN model with 113 parameters.}
    \label{fig:16q_4q_qcnn}
\end{figure}

Now, the question is whether scaling the model leads to improved performance. To investigate this, larger QCNN models with 8-qubits and 16-qubits were implemented and compared to the 4-qubits model and two CNN models with 51 and 113 trainable parameters at threshold 0.9. The 8-qubits QCNN contains 72 trainable parameters and the 16-qubits one has 96 parameters. The results of these models are shown in Fig.~\ref{fig:16q_4q_qcnn}. Despite the increased system size and a higher number of trainable parameters, the 4-qubits QCNN model with HEE encoding still achieves better performance than all other models, and among the QCNN models the HEE 16-qubits one has the worst performance. This suggests that simply increasing the model size does not necessarily lead to better results and may introduce additional complexity that is not effectively utilized. Furthermore, a 16-qubits QCNN model using TPE embedding without entangling gates was also tested and achieved better performance compared to the 16-qubits model with HEE encoding, even though this was the opposite case for both the 4- and 8-qubits models. 

The reduced performance of the 16-qubits QCNN model with HEE encoding raises the question of why increasing the number of trainable parameters does not lead to an improvement, as might be expected. Instead, the performance degradation suggests that scaling the model introduces additional challenges. The differences in performance for the different encoding types highlights the significant role of data embedding in quantum neural networks, suggesting that, for the models considered here, the choice of encoding can have a comparable or even larger impact on performance than increasing the model size. These observations suggest that further improvements may be achieved by optimizing embedding strategies and circuit design rather than relying solely on scaling the number of qubits. Comparing to the CNN model, all QCNN models, except for the HEE 16-qubits system, are performing better. When the CNN model was enlarged to 113 trainable parameters, it showed much worse performance. The test accuracy behavior of this model usually reflects overfitting; however, in this case the training results were very similar to the test results. This suggests that the performance drop may have sources other than overfitting. One possible explanation is that the larger models pick up additional features that are not relevant to the task and may even interfere with identifying the information in the dataset that is most relevant for entanglement classification. Other possibility is the fact that more complex models would have more complex cost function landscapes, which introduces more trainability issues such as falling at local minima, for example. Hence, in the case of the studied dataset here, a small model was enough to achieve optimal performance for the classification task.    

In the case of the entanglement threshold classification task at hand, the model that showed the best accuracy among all models was the 4 qubits HEE QCNN. This model is efficiently classically simulable, which could be considered as a quantum-inspired model. Nevertheless, it outperformed the purely classical model with comparable and larger trainable parameters count. The simplicity of this model is advantageous for the given task in the sense that one could efficiently repeat the learning process for more number of entanglement thresholds and for even shoter ranges in order to accomplish the ultimate goal of determining the entanglement of particle scattering events from fermion scattering density images. 

\vspace{0.5cm}
\section{Conclusion}

In this work, we investigated the feasibility of inferring entanglement properties of fermion scattering processes from easily accessible observables, namely fermion density profiles. By framing the problem as a supervised classification task across multiple entanglement thresholds, we demonstrated that machine learning models can effectively distinguish between different entanglement regimes without requiring direct evaluation of entanglement entropy. This provides a practical route toward extracting nontrivial quantum information from observables that are considerably easier to access.

Specifically, we explored the performance of quantum machine learning architectures, focusing on the QCNN model, and compared them to classical CNNs with comparable parameter counts. Across all tested thresholds, the QCNN consistently achieved competitive or superior classification accuracy, while also exhibiting faster convergence and lower variance. These results suggest that QCNNs can effectively extract information correlated with entanglement regimes from fermion density profiles generated in the scattering process.

An important observation of this study is that increasing model size does not necessarily improve performance. Larger QCNN models with more qubits and parameters exhibited degraded accuracy, despite their higher expressive capacity. This behavior suggests that optimization challenges, increased model complexity, and sensitivity to encoding strategies play a crucial role in determining performance. In particular, the choice of data encoding was found to have a significant impact, in some cases outweighing the effect of scaling the model size. Similarly, enlarging the classical CNN model also led to performance degradation, indicating that the observed behavior is not exclusively quantum but rather reflects a broader interplay between model capacity, data structure, and trainability.

Overall, the best-performing model in this study was the 4-qubits QCNN with HEE encoding, which combines strong accuracy with a relatively small number of parameters and efficient classical simulability. One possible interpretation is that the fermion scattering task considered in this work is simple enough that a compact model already captures the relevant features of the data. Whether larger models become advantageous for more complex scattering processes, such as meson scattering, remains an interesting open question for future study.

In summary, our results provide initial evidence that quantum and quantum-inspired machine learning models can be useful for identifying entanglement regimes from accessible observables. Future work will focus on improving encoding strategies, exploring larger and more diverse datasets, and extending these methods to genuinely quantum data, paving the way toward fully quantum data-driven learning frameworks.

\bibliographystyle{apsrev4-2}
\bibliography{references}

\end{document}